\documentclass[]{article}
\usepackage{lmodern}
\usepackage{amssymb,amsmath}
\usepackage{ifxetex,ifluatex}
\usepackage{fixltx2e} 
\ifnum 0\ifxetex 1\fi\ifluatex 1\fi=0 
  \usepackage[T1]{fontenc}
  \usepackage[utf8]{inputenc}
\else 
  \ifxetex
    \usepackage{mathspec}
  \else
    \usepackage{fontspec}
  \fi
  \defaultfontfeatures{Ligatures=TeX,Scale=MatchLowercase}
\fi
\IfFileExists{upquote.sty}{\usepackage{upquote}}{}
\IfFileExists{microtype.sty}{%
\usepackage{microtype}
\UseMicrotypeSet[protrusion]{basicmath} 
}{}
\usepackage[margin=1in]{geometry}
\usepackage{hyperref}
\hypersetup{unicode=true,
            pdftitle={Artificial life, complex systems and cloud computing: a short review},
            pdfauthor={JJ Merelo},
            pdfborder={0 0 0},
            breaklinks=true}
\usepackage{graphicx,grffile}
\makeatletter
\def\maxwidth{\ifdim\Gin@nat@width>\linewidth\linewidth\else\Gin@nat@width\fi}
\def\maxheight{\ifdim\Gin@nat@height>\textheight\textheight\else\Gin@nat@height\fi}
\makeatother
\setkeys{Gin}{width=\maxwidth,height=\maxheight,keepaspectratio}
\IfFileExists{parskip.sty}{%
\usepackage{parskip}
}{
\setlength{\parindent}{0pt}
\setlength{\parskip}{6pt plus 2pt minus 1pt}
}
\ifx\paragraph\undefined\else
\let\oldparagraph\paragraph
\renewcommand{\paragraph}[1]{\oldparagraph{#1}\mbox{}}
\fi
\ifx\subparagraph\undefined\else
\let\oldsubparagraph\subparagraph
\renewcommand{\subparagraph}[1]{\oldsubparagraph{#1}\mbox{}}
\fi

\let\rmarkdownfootnote\footnote%
\def\footnote{\protect\rmarkdownfootnote}

\usepackage{titling}

  \title{Artificial life, complex systems and cloud computing: a short review}
  \pretitle{\vspace{\droptitle}\centering\huge}
  \posttitle{\par}
  \author{JJ Merelo}
  \preauthor{\centering\large\emph}
  \postauthor{\par}
  \predate{\centering\large\emph}
  \postdate{\par}
  \date{September 2nd, 2017}

\begin{document}
\maketitle
\begin{abstract}
Cloud computing is the prevailing mode of designing, creating and
deploying complex applications nowadays. Its underlying assumptions
include distributed computing, but also new concepts that need to be
incorporated in the different fields. In this short paper we will make a
review of how the world of cloud computing has intersected the complex
systems and artificial life field, and how it has been used as
inspiration for new models or implementation of new and powerful
algorithms
\end{abstract}

\subsection{Introduction}\label{introduction}

\href{https://en.wikipedia.org/wiki/Cloud_computing}{Cloud computing} is
currently the dominant computing platform. Even if initially it was a
metaphor applied to virtualized resources of any kind that could be
accessed in a pay-per-use basis, it has extended itself way beyond its
initial and simple translation of data center concepts to create
completely new software architectures and methodologies for developing,
testing and deploying large scale applications. These new methodologies
have several key issues:

\begin{itemize}
\item
  Infrastructure is fully automatized and described by software; there
  is no \emph{hard} boundary between software and hardware, which are
  developed at the same time, with an application accompanied by the
  description of the resources it needs.
\item
  Applications are a loose collection of resources which interact
  asynchronously, are independent of each other, and in many cases have
  their own vendors or product owners. Some resources are ephemeral,
  appearing when they are needed and vanishing afterwards, some are
  permanent, but all of them behave reactively, acting when they receive
  information, and asynchronously, never waiting for response to return
  but relying instead on promises or other programming language
  constructs.
\item
  As such a collection, cloud-native applications become \emph{organic}
  with parts of it changing without making the application as a whole a
  different one. There is internal evolution with continuous
  integration, testing and deployment, as well as evolution in the sense
  of a progress in fitness, as applications compete with others for mind
  and market share, as well as resources.
\end{itemize}

All these characteristics make the cloud an environment that is closer
to artificial life than single or multidesktop applications. However,
cloud computing is relatively new and, being as it is a techno-social
system (Vespignani 2009,J. J. Merelo et al. (2016)) it will probably
constitute the object of study form the point of view of complex
systems, although for the time being that has not been the case. That is
the reason why we have decided to create this mini-review of how cloud
computing has been applied in the alife/complex systems field, with an
emphasis on those papers written in the last few years, when the cloud
ecosystem has experimented a big diversity expansion, mainly due to the
introduction of containers, isolated applications that act like
lightweight virtual machines. We will do so in the next section.

\subsection{First steps towards cloud as complex
systems}\label{first-steps-towards-cloud-as-complex-systems}

One of the most straightforward ways of connecting \emph{old} algorithms
to \emph{new} technologies is simply to run those algorithms using a
straightforward translation of the old technologies. Cloud computing,
for instance, offers the possibility of instantiating computing nodes at
a cost that is a fraction of buying them; thus it is just
straightforward to port an on premises parallel implementation of
whatever algorithm to a cloud-based one. That kind of transportation
hardly deserves a paper, although it does present some challenges and
has in fact been addressed in several ones; for instance (Medel et al.
2017) studies how to create models of complex systems using cloud
resources and (Merelo-Guervós et al. 2011) presents a system that uses
free cloud storage services as a device for interchanging individuals in
an evolutionary algorithm. Several other artificial life systems that
use the cloud for storage are presented in a recent review (Taylor et
al. 2016); however, there are other ways of doing this kind of
translation. One of the possibilities that cloud computing offers is the
virtualization of computing resources and offering them \emph{as a
service}. For instance, offering robot evolution as a service (Du et al.
2017,Chen, Du, and García-Acosta (2010)) so that anyone can work with
without the need to set up their own infrastructure. Of course, these
new implementations have their own scaling and scheduling issues which
can be a challenge, but they are mostly a straightforward implementation
of \emph{x}-as-a-framework-or-implementation to \emph{x}-as-a-service.
Besides, as shown in the paper mentioned above, virtualizing contributes
to a reduction of costs, which is one of the first-order results of
working in the cloud, but still it is only a short step away from
service-oriented architectures.

This cloud-based setup does have the disadvantage of needing a permanent
connection and possibly a high bandwidth. However, there is an
interesting complex-systems approach to this: so-called \emph{edge
computing} (Satyanarayanan 2017) moves cloud resources \emph{close} to
the service client via technologies that allow them to access other
clients using peer to peer technologies, or using mesh networks or
similar technologies to establish services that users of mainly mobile
devices can access. Since these \emph{edge} nodes kind of
\emph{surround} the user, they receive the denomination of \emph{fog}
computing too (Luan et al. 2015), which is studied mainly in the context
of the Internet of Things. The devices and computing nodes constituting
this \emph{fog} are, effectively, a complex adaptive system (Yan and
Ji-Hong 2010)(Roca et al. 2018). However, the scale and sheer number of
devices used in fog computing exceed by orders of magnitude the one in
actual cloud computing system for the time being, so that for the time
being these concepts are still not being applied to them. It is just,
however, a matter of time and scale when cloud systems will be
considered self-organizing and exhibit emergent behavior. We will refer
to this in the next section.

\subsection{Complex cloud systems}\label{complex-cloud-systems}

Complex cloud systems would make use of the underlying physical (or
virtual) characteristics of the cloud to implement complex systems. Most
cloud systems architectures are based on queues, which usually form the
backbone of the whole system and are used to deliver information as well
as activate different modules in a way that scales well. This feature is
leveraged in (Bottone et al. 2016), which implements an ant-based system
using the MQTT queue management framework to deposit virtual pheromones,
which are then \emph{sniffed} by agents in an stigmergic algorithm.
Instead of simply implementing an array in one of the virtual nodes or a
database which can be read from any client, this is a way that
translates bioinspired algorithms to their closest equivalent in a
virtual environment, creating cloud native complex adaptive systems.

This can be taken further; every module in a cloud system is not
adaptive, performing whatever service it has been programmed to do via
its description. But lately, some cloud systems are being built with
self-\texttt{*} principles in mind; as services in the cloud acquire
autonomy in what is called \emph{autonomic computing}, \emph{adaptation}
is not far away and services become adaptive, being able to
self-organize and the whole system to self-heal, two agencies that make
a cloud system acquire complex adaptive behaviour, such as being able to
reconfigure connections, spawn new copies or eliminate them, all in an
autonomic way without needing to rely on a central authority, using peer
to peer protocols such as the \emph{gossip} protocol (Juan Luís Jiménez
Laredo et al., n.d.), which has actually been used for evolutionary
algorithms (Juan Luis J. Laredo et al. 2009) before cloud computing was
even created.

\subsection{Conclusions}\label{conclusions}

As can be observed by the dearth of references for complex cloud systems
or cloud-native artificial life, we are still in the early stages of its
development, with the cloud being used mainly as a resource for the
straightforward porting of earlier systems. Some steps have been taking
in the realization of the complex adaptive nature of systems and in the
application of complex systems methods such as stigmergy or gossip
protocols to cloud native applications, but it is still an early stage
of development. The next few months or years will undoubtedly bring new
and unexpected developments to this field that will provide benefits in
both directions: insights in the study of cloud as socio-technical
systems and new ways of implementing cloud-native artificial life forms.

\subsubsection{Acknowledgements}\label{acknowledgements}

This work has been supported in part by the Spanish Ministry of Economía
y Competitividad, projects TIN2014-56494-C4-3-P (UGR-EPHEMECH).

\subsection*{References}\label{references}
\addcontentsline{toc}{subsection}{References}

\hypertarget{refs}{}
\hypertarget{ref-bottone2016implementing}{}
Bottone, Michele, Filippo Palumbo, Giuseppe Primiero, Franco Raimondi,
and Richard Stocker. 2016. ``Implementing Virtual Pheromones in Bdi
Robots Using Mqtt and Jason (Short Paper).'' In \emph{Cloud Networking
(Cloudnet), 2016 5th Ieee International Conference on}, 196--99. IEEE.

\hypertarget{ref-chen2010robot}{}
Chen, Yinong, Zhihui Du, and Marcos García-Acosta. 2010. ``Robot as a
Service in Cloud Computing.'' In \emph{Service Oriented System
Engineering (Sose), 2010 Fifth Ieee International Symposium on},
151--58. IEEE.

\hypertarget{ref-du2017robot}{}
Du, Zhihui, Ligang He, Yinong Chen, Yu Xiao, Peng Gao, and Tongzhou
Wang. 2017. ``Robot Cloud: Bridging the Power of Robotics and Cloud
Computing.'' \emph{Future Generation Computer Systems} 74. Elsevier:
337--48.

\hypertarget{ref-laredo09cache}{}
Laredo, Juan Luis J., Carlos Fernandes, Antonio Mora, Pedro A. Castillo,
Pablo Garcia-Sanchez, and Juan Julian Merelo. 2009. ``Studying the Cache
Size in a Gossip-Based Evolutionary Algorithm.'' In \emph{Proceedings of
the 3rd International Symposium on Intelligent Distributed Computing},
edited by G.A. Papadopoulos and C. Badica, 237:131--40. Studies in
Computational Intelligence. Springer-Verlag Berlin Heidelberg.

\hypertarget{ref-LNCS44480129}{}
Laredo, Juan Luís Jiménez, Pedro Angel Castillo, Ben Paechter, Antonio
Miguel Mora, Eva Alfaro-Cid, Anna I. Esparcia-Alcázar, and Juan Julián
Merelo. n.d. ``Empirical Validation of a Gossiping Communication
Mechanism for Parallel EAs.'' In, 129--36.

\hypertarget{ref-luan2015fog}{}
Luan, Tom H, Longxiang Gao, Zhi Li, Yang Xiang, Guiyi Wei, and Limin
Sun. 2015. ``Fog Computing: Focusing on Mobile Users at the Edge.''
\emph{arXiv Preprint arXiv:1502.01815}.

\hypertarget{ref-Medel2017}{}
Medel, Victor, Unai Arronategui, José Ángel Bañares, and José-Manuel
Colom. 2017. ``Distributed Simulation of Complex and Scalable Systems:
From Models to the Cloud.'' In \emph{Economics of Grids, Clouds,
Systems, and Services: 13th International Conference, Gecon 2016,
Athens, Greece, September 20-22, 2016, Revised Selected Papers}, edited
by José Ángel Bañares, Konstantinos Tserpes, and Jörn Altmann, 304--18.
Cham: Springer International Publishing.
doi:\href{https://doi.org/10.1007/978-3-319-61920-0_22}{10.1007/978-3-319-61920-0\_22}.

\hypertarget{ref-JJ2016}{}
Merelo, J. J., Paloma de Las Cuevas, Pablo García-Sánchez, and Mario
García-Valdez. 2016. ``The Human in the Loop: Volunteer-Based
Metacomputers as a Socio-Technical System.'' In \emph{Proceedings of the
Artificial Life Conference 2016}, 648. Complex Adaptive Systems. One
Rogers Street Cambridge MA 02142-1209: The MIT Press.
doi:\href{https://doi.org/http://dx.doi.org/10.7551/978-0-262-33936-0-ch103}{http://dx.doi.org/10.7551/978-0-262-33936-0-ch103}.

\hypertarget{ref-merelo2011evostar}{}
Merelo-Guervós, Juan-Julián, Maribel García Arenas, Antonio Miguel Mora,
Pedro A. Castillo, Gustavo Romero, and Juan Luís Jiménez Laredo. 2011.
``Cloud-Based Evolutionary Algorithms: An Algorithmic Study.''
\emph{CoRR} abs/1105.6205.

\hypertarget{ref-roca2018tackling}{}
Roca, Damian, Rodolfo Milito, Mario Nemirovsky, and Mateo Valero. 2018.
``Tackling Iot Ultra Large Scale Systems: Fog Computing in Support of
Hierarchical Emergent Behaviors.'' In \emph{Fog Computing in the
Internet of Things}, 33--48. Springer.

\hypertarget{ref-satyanarayanan2017edge}{}
Satyanarayanan, Mahadev. 2017. ``Edge Computing: Vision and
Challenges.'' USENIX Association.

\hypertarget{ref-taylor2016webal}{}
Taylor, Tim, Joshua E Auerbach, Josh Bongard, Jeff Clune, Simon
Hickinbotham, Charles Ofria, Mizuki Oka, Sebastian Risi, Kenneth O
Stanley, and Jason Yosinski. 2016. ``WebAL Comes of Age: A Review of the
First 21 Years of Artificial Life on the Web.'' \emph{Artificial Life}.
MIT Press.

\hypertarget{ref-vespignani2009predicting}{}
Vespignani, Alessandro. 2009. ``Predicting the Behavior of Techno-Social
Systems.'' \emph{Science} 325 (5939). American Association for the
Advancement of Science: 425--28.

\hypertarget{ref-yan2010application}{}
Yan, Chen, and Qiao Ji-Hong. 2010. ``Application Analysis of Complex
Adaptive Systems for Wsn.'' In \emph{Computer Application and System
Modeling (ICCASM), 2010 International Conference on}, 7:V7--328. IEEE.

\end{document}